\documentclass[preprint]{aastex}

\usepackage{epsf}

\def\linebreak{\hfil\break}

\def\
{\hfil\linebreak}

\def\lstar{\ifmmode {L_{\star}}\else $L_{\star}$\fi}
\def\tstar{\ifmmode {T_{\star}}\else $T_{\star}$\fi}
\def\msun{\ifmmode {\rm M_{\odot}}\else $\rm M_{\odot}$\fi}
\def\MSUN{\ifmmode {\rm M_{\odot}}\else $\rm M_{\odot}$\fi}
\def\msunyr{\ifmmode {\rm M_{\odot}\,yr^{-1}}\else $\rm M_{\odot}\,yr^{-1}$\fi}
\def\mdot{\ifmmode {\dot{M}}\else $\dot{M}$\fi}
\def\lsun{\ifmmode {\rm L_{\odot}}\else $\rm L_{\odot}$\fi}
\def\LSUN{\ifmmode {\rm L_{\odot}}\else $\rm L_{\odot}$\fi}
\def\lbol{\ifmmode {L_{bol}}\else $L_{bol}$\fi}
\def\teff{\ifmmode {T_{eff}}\else $T_{eff}$\fi}
\def\ne{\ifmmode {n_{e}}\else $n_{e}$\fi}
\def\te{\ifmmode {T_{e}}\else $T_{e}$\fi}
\def\rc{\ifmmode {R_{c}}\else $R_{c}$\fi}
\def\degree{\ifmmode {^\circ}\else {$^\circ$}\fi}
\def\mum{\ifmmode {\rm \mu {\rm m}}\else $\rm \mu {\rm m}$\fi}
\def\arcsec{\ifmmode ^{\prime \prime}\else $^{\prime \prime}$\fi}
\def\secpoint{\mbox{$''\mskip-7.6mu.\,$}}

\def\inch{\ifmmode ^{\prime \prime}\else $^{\prime \prime}$\fi}
\def\arcmin{\ifmmode ^{\prime}\else $^{\prime}$\fi}

\def\msun{\ifmmode {\rm M_{\odot}}\else $\rm M_{\odot}$\fi}
\def\mearth{\ifmmode {\rm M_{+\mskip-14.6muO\,}}\else $\rm M_{+\mskip-14.6muO\,}$\fi}
\def\mearth{\ifmmode {\rm M_{\earth}}\else $\rm M_{\earth}$\fi}
\newbox\grsign \setbox\grsign=\hbox{$>$} \newdimen\grdimen \grdimen=\ht\grsign
\newbox\simlessbox \newbox\simgreatbox
\setbox\simgreatbox=\hbox{\raise.5ex\hbox{$>$}\llap
     {\lower.5ex\hbox{$\sim$}}}\ht1=\grdimen\dp1=0pt
\setbox\simlessbox=\hbox{\raise.5ex\hbox{$<$}\llap
     {\lower.5ex\hbox{$\sim$}}}\ht2=\grdimen\dp2=0pt

\def\simless{\mathrel{\copy\simlessbox}}
\shorttitle{AG Pegasi}
\shortauthors{KENYON, PROGA, and KEYES}
\begin{document}

\title{The Continuing Slow Decline of AG Pegasi}

\author{Scott J. Kenyon}
\affil{Smithsonian Astrophysical Observatory,
60 Garden Street, Cambridge, MA 02138}
\affil{skenyon@cfa.harvard.edu}

\author{Daniel Proga}
\affil{NASA Goddard Space Flight Center,
Laboratory for High Energy Astrophysics,
Code 662, Greenbelt, MD 20771}
\affil{proga@sobolev.gsfc.nasa.gov}

\and

\author{Charles D. Keyes}
\affil{Space Telescope Science Institute,
3700 San Martin Drive, Baltimore, MD 21218}
\affil{keyes@stsci.edu}
\affil{$  $}
\affil{\it to be published in the}
\affil{\bf Astronomical Journal}
\affil{\it July 2001}

\begin{abstract}

We analyze optical and ultraviolet observations of the symbiotic binary 
AG Pegasi acquired during 1992--97.  The bolometric luminosity of the 
hot component declined by a factor of 2--3 from 1980--1985 to 1997.  
Since 1992, the effective temperature of the hot component may have declined 
by 10\%--20\%, but this decline is comparable to the measurement errors. 
Optical observations of H$\beta$ and He~I emission show a clear illumination 
effect, where high energy photons from the hot component ionize the outer
atmosphere of the red giant.  Simple illumination models generally account 
for the magnitude of the optical and ultraviolet emission line fluxes.  
High ionization emission lines -- [Ne~V], [Mg~V], and [Fe~VII] -- suggest 
mechanical heating in the outer portions of the photoionized red giant wind.  
This emission probably originates in a low density region $\sim$
30--300 AU from the central binary.  

\end{abstract}

\subjectheadings{stars: evolution -- stars: novae -- stars: 
individual (AG Pegasi)}

\section{INTRODUCTION}

The symbiotic binary AG Peg began a slow nova eruption in the mid-1850's,
rising from 9th magnitude to 6th magnitude in $\sim$ one decade or less
\citep{lun21}.  Although spectra at optical maximum are not available, 
later spectra revealed a peculiar Be-type spectrum with strong P Cygni-like 
emission lines and additional absorption from He I and TiO \citep[][1929a,b,
1959]{fle07, cpg57, mer16}.  Since the 1920's, the hot component 
of the binary has slowly evolved from a Be-type star to a Wolf-Rayet star 
(WN6 spectrum) to a hot ``subdwarf'' with an effective temperature 
exceeding $10^5$ K \citep[][and references therein]{gal79,ken93}.
The red giant companion to the eruptive star has an M3 spectral type, 
does not fill its tidal lobe, and loses mass in a low velocity wind 
\citep{ken87,kny91,mur99}.

The eruption of the hot component in AG Peg may be the slowest classical 
nova outburst ever recorded \citep{gal79,ken83,nus92,ken93}. Historical
data -- together with ultraviolet (UV) spectra acquired with the 
{\it International Ultraviolet Explorer (IUE)} and other UV satellite 
missions -- indicate that the hot component maintained a roughly constant 
bolometric luminosity from $\sim$ 1850 to $\sim$ 1980.  The last few pairs 
of {\it IUE} spectra suggest that the luminosity of the hot component began 
to decline as the intensity of the broad emission lines weakened considerably 
\citep[][Nussbaumer, Schmutz, \& Vogel 1995]{ken93,vog94}.  At about the same 
time, the source of strong emission lines may have shifted from an H~II region 
surrounding the hot component to the illuminated hemisphere of the 
red giant (e.g., Proga, Kenyon, \& Raymond 1998).  Radio, X-ray,
and other satellite data show features indicative of colliding winds, 
where material ejected from the hot component interacts with the 
slow-moving wind of the red giant \citep{kny91,mur95,nus95,con97}.  
If these analyses are correct, the hot component may be slowly 
transforming from a slow classical nova at maximum into a nova remnant.

To elucidate further the slow decline of the hot component in AG Peg,
we analyzed new optical spectra and UV spectra acquired with the
{\it Hubble Space Telescope (HST)}.  The optical spectra demonstrate
that most of the low ionization emission lines, such as H~I and He~I,  
arise in the photoionized wind of the red giant.  Some high ionization
emission from He~II is also produced in this region.  Although more
spectra are required to follow the evolution of the hot component in
detail, the UV data provide further evidence for a decline in the 
bolometric luminosity of the hot component.  The forbidden and
intercombination emission lines indicate that an ionized nebula 
with a wide range of densities, $\sim$ $10^5$ to $10^{10}$ cm$^{-3}$,
surrounds the binary system.  Lines of low ionization potential,
e.g., [O~III], suggest a low density, photoionized nebula with 
an electron temperature, $T_e \sim 10^4$ K; lines of higher 
ionization potential, e.g., [Ne~V] and [Fe~VII], are formed at
higher electron density, $n_e \sim 10^7$ cm$^{-3}$, in gas 
mechanically heated to electron temperatures of at least $10^5$ K. 

We describe the {\it HST} and new optical data in \S2, analyze these
data in \S3, and conclude with a brief summary in \S4.

\section{OBSERVATIONS}

We acquired high quality ultraviolet and optical spectra of AG Peg
with the {\it Faint Object Spectrograph} (FOS) on board the 
{\it Hubble Space Telescope}.  These data were taken as part of 
a calibration program to evaluate the absolute FOS wavelength scale
and to measure the throughput for the small apertures and the amount 
of internal ``out of band'' grating scatter recorded in FOS exposures. 
The spectra cover 1150--4800 \AA~at a resolution of R = 1300 over
the entire spectrum.
To attain the best photometric accuracy, the most precise target 
acquisition and centering procedure was used.  The centering accuracy 
should be roughly 0\secpoint025 for all of the data, with a 1$\sigma$
uncertainty in the HST guiding and orbit-to-orbit re-centering of 
0\secpoint007.  Most of the observations were taken in RAPID mode, which 
yields several individual readouts of the spectrum.  We averaged 
all readouts to produce a final spectrum for each wavelength region.
We acquired observations with both FOS detectors, FOS/BL and FOS/RD,
through several apertures, 4\secpoint3 square, 1\secpoint0 round, 
0\secpoint5 round, and 0\secpoint3 round.  Observations with the largest
apertures should have the best photometric accuracy. In practice, line
fluxes derived from different apertures agreed to 5\% or better; we
thus averaged the line and continuum fluxes described below.
	
We reduced the data using the standard ``calfos'' pipeline routines 
in STSDAS using the STScI-recommended reference files and calibration 
procedures appropriate to each epoch of observation.  Thus, the FOS 
calibration closeout reference file suite, as described in the HST Data 
Handbook, Volume II \citep{key97}, was used for all data. 

P. Berlind, D. Koranyi, and T. Lappin acquired low resolution optical 
spectra of AG Peg with the FAST spectrograph mounted at the Fred L. 
Whipple Observatory 1.5-m telescope on Mount Hopkins, Arizona.
FAST is a high throughput, slit spectrograph equipped with 
a thinned, back-side illuminated 320 $\times$ 2704 Loral CCD  
\citep{fab98}.  These spectra cover 3800--7500 \AA~at a resolution 
of $\sim$ 3 \AA~with a 3\arcsec~slit.
We calibrated the FAST spectra on the \citet{hay75} flux scale
using observations of standards from \citet{bar82}.
Repeat observations of several standards and a comparison
of photoelectric B and V magnitudes with those derived from 
the spectra indicate the calibration has an uncertainty of
$\pm$0.10 mag.

We acquired additional optical spectrophotometry of AG Peg on 
1992 November 5--11 with the white spectrograph and GoldCam CCD 
dewar (WhiteCam) at the Kitt Peak National Observatory (KPNO) 
0.9-m telescope.  Grating 26 and a 7\secpoint5 slit resulted in 
a resolution of $\sim$ 10 \AA~on a Ford 3K $\times$ 1K CCD.
Roughly 200--300 pixels at each end of the CCD are vignetted 
with WhiteCam, which reduced our free spectral range to
3500--6200 \AA~at 1.25 \AA/pixel.  This interval includes 
most of the strong emission lines detected on the FAST spectra
but misses H$\alpha$ and several He~I lines.
We reduced these data to the \citet{hay75} flux scale using 
observations of standard stars from \citet{bar82} and estimate
our flux calibration has an uncertainty of $\pm$0.05 mag.

Figure 1 shows a composite {\it FOS} spectrum.  The data shown in 
this figure and all subsequent figures have not been corrected for
interstellar extinction. The spectrum has
a Rayleigh-Jeans tail from a hot continuum for $\lambda \lesssim
2000$ \AA, a flat nebular continuum for $\lambda \approx$ 2000--3646 \AA,
and a red continuum from the cool giant for $\lambda \gtrsim$ 4000 \AA.
Strong emission lines from a wide range of ionization states are
superimposed on these continua.  FAST spectra show a red continuum,
strong TiO absorption bands, and more emission lines at longer
wavelengths \citep[see Figure 1 of][]{ken93}.  The overall appearance
of the spectrum has changed slightly since the early 1990's: the 
Rayleigh-Jeans tail and the nebular continuum are more prominent 
on the {\it FOS} spectra than on earlier {\it IUE} spectra.

To quantify the spectral changes of AG Peg, we measured continuum
magnitudes and emission line fluxes on all of the spectra.  We define
narrow-band continuum magnitudes in 30 \AA~bandpasses as

\begin{equation}
m_{\lambda} = {\rm -2.5 ~ log} ~ F_{\lambda} - 21.1 ~ ,
\end{equation} 

\noindent
where $F_{\lambda}$ is the average flux in the bandpass. 
Tables 1 and 2 list continuum magnitudes and 1$\sigma$ errors
derived from the {\it FOS} and the FAST/WhiteCam observations.
We measured emission line fluxes using interactive gaussian
fitting routines.  Simple gaussians produce good fits for all
emission lines except H$\alpha$, which has a broad wing superimposed
on a narrow gaussian feature. We used two gaussians to fit this
feature.  Tables 3--5 list emission line fluxes for the
{\it FOS} and the FAST/WhiteCam observations.  The typical error
for FAST measurements in Table 5 is 5\% for strong lines and 
10\% for weak lines based on repeat FAST observations acquired 
on the same night and a comparison of contemporaneous FAST 
and {\it FOS} observations.  The {\it FOS} line fluxes in 
Tables 3--4 have accuracies of 5\% or better.

Figure 2 shows the continuing slow decline of the far ultraviolet
(far-UV) and near-ultraviolet (near-UV) continuum.  We adopt the
\citet{fer85} photometric ephemeris to time maximum light in the 
system, when the hot component lies in front of the red giant:

\begin{equation}
{\rm Max(V)} ~ = ~ {\rm JD} ~ 2442710.1 + 816.5 \cdot {\rm E} .
\end{equation}

\noindent  
Data in Figure 2 and subsequent figures for photometric phases less than 7
are from \citet{key81} and \citet{ken93}.
In addition to the 2.0--2.5 mag decline 
at 1300--3300 \AA, there are obvious maxima in the near-UV continuum 
at $\phi$ = 2, 7, and 8.  These maxima have amplitudes of 0.5-1.0 mag
and are weak or absent in the far-UV.

Figures 3 and 4 show that several emission lines vary in step with
the near-UV continuum.  The N~IV] $\lambda$1486 line rises by a factor
of two at $\phi$ = 2 and has a weaker rise at $\phi$ = 8.  He II 
$\lambda$1640 displays the opposite behavior, with a weak rise at $\phi$ = 2
followed by a stronger variation at $\phi$ = 8.  Both lines were
saturated on {\it IUE} spectra at $\phi$ = 7.  The optical lines
shown in Figure 4 have more obvious variations correlated with 
photometric phase.  The H~I and He~I lines are roughly a factor of two
more intense at $\phi$ = 8 and $\phi$ = 9 than at photometric minima.
Higher ionization optical lines, such as He II $\lambda$4686, have
a weak maximum.  Both sets of lines are a factor of 2--3 weaker 
at $\phi$ = 8--9 compared to data at $\phi$ = 2--4.

\section{ANALYSIS}

The behavior of the lines and continuum in AG Peg continue to support 
the phenomenological model developed by \citet[][b, 1970]{bl68a}, 
\citet[][1967]{boy66}, and \citet{gal79}.  The overall 
decrease in the optical and UV
line and continuum fluxes results from the continuing slow nova-like
decline of the hot component.  \citet{ken93} showed that the
bolometric luminosity of the hot component has remained roughly
constant for nearly a century \citep[see also][]{gal79}.  
As the photospheric radius contracted, the effective temperature
of the hot component increased. The optical and UV continua faded.  
Fluxes from low ionization lines also declined;
fluxes from the highest ionization lines increased. Because high 
energy photons from the hot component ionize the outer atmosphere
of the red giant, the system displays photometric maxima in phase 
with the orbit.  The nebular continuum and recombination lines of 
H~I and He~I are strongest when we see the heated hemisphere and 
weakest when we see the opposite, unheated hemisphere.  Prior to
1990, higher ionization lines formed closer to the hot component; 
the fluxes of these lines thus were only weakly modulated by the orbit.

The high quality {\it FOS} and FAST data yield new tests of this picture. 
Based on an analysis of {\it IUE} data, \citet{ken93} suggested
that AG Peg might be evolving towards lower luminosities along a white 
dwarf cooling curve.  They proposed that the transition from a Wolf-Rayet
spectrum with broad emission lines to a nebular spectrum with narrow emission
lines marked the start of this evolution.  At the same time, the source
of the strong high ionization emission lines moved from an H~II region
surrounding the hot component to the illuminated hemisphere of the red
giant.  \citet{con97} and several other studies suggest that some emission
features form in an interaction region between the two stars, where the
low velocity wind of the red giant collides with the higher velocity wind
from the hot component.

To test these proposals in more detail, we first consider the evolution of
the hot component in \S3.1.  We then analyze emission lines produced in
the illuminated red giant atmosphere (\S3.2) and in shocked gas in the
outer portions of the red giant wind (\S3.3). We conclude this section
with some comments on colliding winds in the AG Peg binary (\S3.4).

\subsection{Evolution of the Hot Component}

To understand the evolution of the hot component, we follow previous studies 
and derive the time variation of luminosity $L_h$ and effective temperature
$T_h$.  Because the complete energy distribution of the hot component cannot
be observed directly, we need accurate proxies for $L_h$ and $T_h$.  Three
techniques have been proposed in previous studies. \citet{kw84} adopted 
a blackbody model for the hot component and predicted UV color indices 
for sources surrounded by a photoionized nebula; \citet{ken85} generalized 
this approach and derived $L_h$ and $T_h$ from fits to the UV spectral energy 
distribution \citep[see also][]{ken93}.  
\citet{fer88} developed a UV Zanstra temperature to derive $T_h$ from 
measurements of the He~II $\lambda$1640 emission line and the continuum 
flux at 1400 \AA.  M\"urset et al. \citep[1991; see also][]{key81,mur94} 
estimated $L_h$ and $T_h$ from He~II $\lambda$1640 and the UV spectral 
energy distribution.  Following \citet{iij81}, \citet{ken86} inferred 
$L_h$ and $T_h$ from the optical H~I and He~II lines, which are less 
sensitive to uncertainties in the interstellar extinction to the source.

Figure 5 shows results for the time variation of $L_h$ and $T_h$ in AG Peg
from these three techniques.  We used data from \citet{key81}, \citet{ken93},
and this
paper, corrected for $E_{B-V}$ = 0.10 using the \citet{mat90} extinction
curve for an adopted distance of 800 pc.  We estimate $L_h$ and $T_h$ from
(i) the UV continuum fits of \citet{ken85},
(ii) the UV Zanstra temperature of \citet{fer88}, and
(iii) the optical Zanstra temperature of \citet{ken86}.
The formal uncertainties are $\pm$ 0.1 in log $L_h$ and 
$\pm$ 0.05 in log $T_h$.  The uncertain distance adds a systematic 
uncertainty to the luminosity but not to the effective temperature.
In the top panel, the luminosity derived from the UV Zanstra
temperature monotonically decreases from $\phi$ = 1 to $\phi$ = 9 (filled
circles).  The luminosity derived from the UV continuum is roughly constant
with time and may decline some at $\phi \ge$ 7--8 (plus signs).  As expected,
the luminosity derived from optical data follows the illumination effect
(open circles):
$L_h$ is largest at photometric maxima and smallest at photometric minima.
The average luminosity is comparable to the $L_h$ derived from UV data
and shows the same overall trend: a roughly constant $L_h$ for $\phi$ = 1--7
followed by a decline.

The time evolution of $T_h$ is shown in the lower panel of Figure 5. The $T_h$ 
derived from the optical emission lines is always 25\% to 50\% larger than 
the $T_h$ estimated
from the UV lines.  This difference is easily understood.  Because the H~I
line fluxes change by factors of 2--3 with orbital phase, the flux ratio
F($\lambda$4686)/F($\lambda$4861) is largest at photometric minima and 
smallest at photometric
maxima.  The $T_h$ derived from this ratio thus varies with phase, as shown
by the open circles in Figure 5.  The $L_h$ derived from the optical line 
fluxes is anti-correlated with $T_h$.  For much of the evolution, the $T_h$
derived from the UV continuum is 25\% to 50\% smaller than the $T_h$ estimated
from the UV lines (plus signs).  AG Peg resembled a Wolf-Rayet star during this 
period; its UV energy distribution was very similar to the energy distribution
of the luminous Wolf-Rayet star HD 50896.  Assuming that the \citet{hil87}
models for HD 50896 also fit AG Peg, \citet{ken93} derived a luminosity
for the hot component in AG Peg from the observed {\it Voyager} spectrum.
This estimate is $\sim$ 50\% larger than derived from UV continuum fits.
To produce a larger luminosity from the same continuum flux, the $T_h$ 
implied by this analysis must also be $\sim$ 50\% larger.  Broad Wolf-Rayet 
emission lines began to disappear from low resolution UV spectra of AG Peg 
at $\phi \ge$ 6.  The UV continuum data then begin to yield a $T_h$ similar 
to that derived from the UV Zanstra temperature. 

This analysis suggests that the results for $L_h$ and $T_h$ from the UV Zanstra
temperature are more reliable indicators of the evolution of the hot component
in AG Peg. We conclude that $L_h$ was roughly constant during $\phi$ = 1--3,
when $T_h$ rose by $\sim$ 20\%.  As $T_h$ remained roughly constant at $\phi$
= 3--8, $L_h$ declined by a factor of 2--3.  The {\it HST} data suggest that
$T_h$ may have declined by $\sim$ 10\% to 20\% during $\phi$ = 8.0--9.5, but
this decline is comparable to the size of the uncertainties in the temperature
estimates.

\subsection{Illumination of the Red Giant}

Since the 1930's, the ionization of the red giant atmosphere by the hot 
component has been an important feature of the AG Peg binary 
\citep[see][and references therein]{mer59,bel70}.  Merrill's
observations demonstrated that fluxes of low ionization emission lines 
vary in phase with the orbital motion of the red giant.  \citet{bel70} 
interpreted these variations as the reflection effect, where the hot 
component heats the outer red giant atmosphere.  She showed that the 
observed amplitude and the relative phasing of the variation were 
consistent with model predictions.  The observed variations in He~II
and other high ionization lines, however, were not consistent with the
reflection effect. The radial velocity variations and the broad widths -- 
$\sim$ 1000 km s$^{-1}$ -- of these features were more consistent with line
formation in the outer atmosphere of the hot component.

Evolution of the emission line fluxes in 1985--1990 suggest a modest 
change in this picture.  The broad WN-type emission lines from He~II,
N~V, and other species faded to reveal narrow emission features with
profiles similar to those of low ionization lines (Kenyon et al. 1993).
Although the phase coverage of the UV data is poor after $\sim$ 1990,
our optical data show a weak modulation of He~II with orbital phase
(Figure 4).  This behavior suggests that at least some He~II emission
is produced in the ionized red giant atmosphere.

To investigate the illumination of the red giant atmosphere
in more detail, \citet[][1998]{pro96} developed a non-LTE 
photoionization model.  They showed that the spectrum of an 
illuminated red giant atmosphere is too weak to explain the strong 
line spectrum observed in AG Peg and other symbiotic stars.  A normal 
red giant does not intercept enough radiation from the hot component 
to produce prominent emission lines \citep{pro96}.  However, 
a red giant wind can intercept enough high energy photons from the
hot component if the optical depth in the wind is large \citep{pro98}.
Winds with mass loss rates of $10^{-7}$ \msunyr~can generally account
for the observations of symbiotics such as AG Peg.  These models
reproduce observed fluxes of metallic lines if the O/N and C/N abundances
are similar to those of field red giants.  Predicted fluxes for H and
He lines fall below the observations.  

\citet{pro98} summarized the simplifications and weaknesses of
their calculations.  They derived an accurate solution for illumination
along the line of centers connecting the two stars and assumed that the
total flux from the illuminated atmosphere $F_{tot}$ is

\begin{equation}
F_{tot} = r_{cs} F_{1D} ~ ,
\end{equation}

\noindent
where $F_{1D}$ is the flux derived from a unit area along the line of 
centers and $r_{cs}$ is the geometric cross-section of the red giant wind.
This simplification ignores any wind from the hot component,
the complex geometry of the red giant wind, and any interaction
region between the two winds.  Proga et al. noted that this simplified
approach was a first step in understanding complex systems like AG Peg
and many other symbiotics.  They suggested several ways to reconcile 
the model with the observations and commented that better observations 
of weaker UV and optical lines might allow a choice between different models.

As a further guide to understanding illumination in AG Peg, we compare
our new data with the theoretical results of \citet[][1998]{pro96}.
Figure~6 indicates line fluxes for wind models and observations of AG~Peg
(the lines are $H\beta$, He~I~$\lambda$5876, O~III]~$\lambda$1664,
CIV~$\lambda$1550, He~II~$\lambda$4686, and N~V~$\lambda$1240).
We use the data from this paper (filled triangles, $\phi \approx$ 9)
and from \citet[][open circles, $\phi \approx$ 7--8]{ken93}.  The 
solid and dashed lines indicate model predictions of \citet[][1998]{pro96}.
We adopt $T_h = 10^5$~K and $L_h=620~\LSUN$ for the
hot component, and use red giant parameters for Vogel's velocity law 
\citep{vog91} with red giant mass loss rates of
$\mdot = 10^{-8}$ and $10^{-6}~\msunyr$
\citep[see also Figure~12 in][]{pro98}.  Solid curves show predictions 
for $r_{cs} = R_g^2$; dashed curves show predictions for $r_{cs} = 3^2 R_g^2$.
The line fluxes increase with \mdot~for each set of models.  Because 
the emission measure of the ionized wind increases linearly with \mdot,
the predicted line fluxes vary approximately linearly with \mdot~at high 
\mdot. At lower mass loss rates, the line fluxes are limited by the 
amount of material in the {\it static} red giant atmosphere.  This 
material provides a lower limit to the emission measure -- for a 
particular hot component luminosity -- so the line fluxes vary 
little for \mdot~$\simless$ $10^{-8}~\msunyr$.
 
Figure~6 shows that the fluxes of all emission lines decreased by a factor 
of 1.5--3.0 during $\phi = 7-9.5$.  This decline in the fluxes coincides
with the appearance of a modest orbital modulation of the He II line
fluxes (see Figures 3--4).  We interpret this behavior as a change in
the location of high ionization emission lines. Prior to $\phi$ = 7, nearly
all of the high ionization lines formed in a compact H~II region 
surrounding the hot component (see Figure 11 of Kenyon et al. 1993).  
These lines were broad and diffuse, as expected for
material in an outflowing wind from a hot, luminous white dwarf star.
After $\phi$ = 7, the lines are narrow; the illuminated red giant atmosphere
produces some of the high ionization line emission.  Thus, the illumination 
model can better account for observations at $\phi \approx 9$ than at 
$\phi \approx 7$.  Observed fluxes at $\phi$ = 7 lie above optimistic
model predictions (dashed lines); data for $\phi$ = 9 generally lie below
these predictions.  Although the illumination models are still crude,
it is encouraging that as the wind from the hot component weakens and 
AG~Peg becomes a less complex system, the models can better explain 
the observations with reasonable input parameters.
 
\subsection{High Ionization Forbidden Lines}

The ionized nebula surrounding AG Peg consists of several distinct 
components.  The 1--20 cm VLA data are consistent with an unresolved 
central source embedded in several extended shells of gas 
\citep{kny91,sea92,tay84}.  The unresolved radio source is 
probably optically thick, free-free emission from the photoionized
wind of the red giant \citep{kny91,sea92,ken93}. \citet{ken93}
suggested that most of the [O~III] and [Ne~III] emission observed in 
1980--93 is also produced in this volume \citep[see also][]{kny91}.  
These lines have continued
to decline in intensity since 1993; their flux ratios are consistent
with an electron density, $n_e \simless 10^{7}$ cm$^{-3}$, for an
electron temperature, $T_e \sim 10^4$ K \citep{fer78}.
This evolution suggests an overall decrease in the electron density
during the past decade.

\citet{ken93} noted that results for high ionization forbidden lines 
were uncertain because the lines were not very prominent on their
relatively low signal-to-noise spectra.  The new FAST and {\it HST} 
data allow us to make more progress on this region.  We have accurate 
UV and optical fluxes for several [Fe~VII], [Mg~V], and [Ne~V] lines, 
which yield more information on the ionized nebula.

We begin with the [Fe~VII] lines, which have remained roughly constant
in intensity since their first appearance in 1984--1986.  Several 
[Fe VII] intensity ratios, $I(\lambda$5721)/$I(\lambda$6087) and
$I(\lambda$3586)/$I(\lambda$3758), do not depend on $n_e$ or $T_e$.
Our result for $I(\lambda$5721)/$I(\lambda$6087), 0.6--0.8,
brackets the predicted value of 0.65; our result for
$I(\lambda$3586)/$I(\lambda$3758), 0.3, is significantly smaller 
than the predicted value of 0.75.  The small observed value of 
the $I(\lambda$3586)/$I(\lambda$3758) ratio suggests that the 
$\lambda$3758 line is blended with an O~III Bowen fluorescence
line.  The relative intensities of other O~III lines suggest that O~III
$\lambda$3758 should contribute $\sim$ 2/3 to 3/4 of the $\lambda$3758
flux.  If so, the [Fe~VII] $I(\lambda$3586)/$I(\lambda$3758) ratio is 
then close to the predicted value \citep[see Tables 3--4 and][]{sar80}.

With this correction to the $I(\lambda$3758) flux, the reddening-corrected 
[Fe~VII] intensity ratios sensitive to density and temperature are 
$I(\lambda$2015)/$I(\lambda$3758) $\approx$ 0.4--0.6 and
$I(\lambda$3758)/$I(\lambda$6087) $\approx$ 1.1--1.4.
Both intensity ratios are consistent with
$T_e \lesssim 3 \times 10^4$~K if $n_e \gtrsim 10^8$~cm$^{-3}$ and
$T_e \gtrsim 5 \times 10^4$~K if $n_e \lesssim 10^8$~cm$^{-3}$
\citep[][1991]{nus82,kee87}.  Weak detections of [Fe~VII] 
$\lambda \lambda$4942, 5159 favor $n_e \sim 10^7$ cm$^{-3}$
over other values.

To better constrain the physical conditions in the $\rm Fe^{+6}$ region,
we also consider data for [Ne~V]. The ionization potential of $\rm Ne^{+4}$
(126~eV) is comparable to  $\rm Fe^{+6}$ (128~eV); [Ne V] and [Fe VII] 
should form in similar physical conditions.
The intensity ratio for $I(\lambda$3346)/$I(\lambda$3426) is 0.3--0.5,
which brackets the predicted value of  0.36 \citep{nus79}.
The $I(\lambda$3426)/$I(\lambda$2976) intensity ratio is sensitive 
to $n_e$ and $T_e$ \citep{kaf80}.  Our measured ratio of 
$I(\lambda 3426)/I(\lambda2976)$ $ \approx $ 17 requires 
$n_e \lesssim 10^7~{\rm cm^{-3}}$ for $T_e \gtrsim 10^5$~K. 

The critical densities for all of the [Fe~VII] and [Ne~V] lines on our 
spectra are $10^7~{\rm cm^{-3}}$ to $10^8~{\rm cm^{-3}}$ 
\citep[][1991]{kaf80,nus82,kee87}.  
Although there are some uncertainties in the 
atomic physics for both ions, the line ratios strongly imply line
formation in a very hot, low density gas.  This result favors mechanical
heating over photoionization as the energy source for this emission.
To place a better limit on the electron density and on the origin of the
highly ionized forbidden lines, we can estimate the forbidden line fluxes
expected from a photoionized gas.  The volume $V$ for the [Fe~VII]
Str\"omgren sphere is 
\begin{equation}
V = N_{\gamma} / n_{Fe} n_e \alpha_r,
\end{equation}
\noindent
where 
$N_{\gamma}$ is the number of $Fe^{+5}$-ionizing photons,
$n_{Fe}$ is the number density of Fe atoms, and
$\alpha_r$ is the recombination rate.
The luminosity in a single [Fe~VII] emission line is
\begin{equation}
L_{[Fe~VII]} = x_i ~ n_{Fe^{+6}} ~ A_{ji} ~ h \nu_{ij} ~ V ~ ,
\end{equation}
\noindent
where
$x_i$ is the fraction of Fe VII ions in the $j$th level,
$n_{Fe^{+6}}$ is the number density of Fe~VII ions,
$A_{ji}$ is the transition probability, and
$\nu_{ij}$ is the line frequency.
If we assume that all of the Fe in the nebula is in the form of Fe VII,
substitute our expression for the volume into equation (5) and adopt 
the appropriate coefficients \citep{woo81,nus82,kee87,arn92}, the 
luminosity in the $\lambda$6087 line is 
\begin{equation}
L_{\lambda 6087} \approx 0.1 ~ \lsun ~ \left ( \frac{N_{\gamma}}{10^{43} ~ \rm s^{-1}} \right ) \left ( \frac{10^8 ~ \rm cm^{-3}}{n_e} \right ) ~ .
\end{equation}
\noindent
The coefficient in equation (6) varies by a factor of 2--3 for $T_e$ = 
$10^4$ K to $10^5$ K if the hot component has an effective temperature
of $10^5$ K and emits as a blackbody.  This result -- together with 
similar expressions for other [Fe~VII] and [Ne~V] lines -- suggests 
that photoionization can account for the observed optical and ultraviolet 
line fluxes, $L_{obs} \sim$ 0.05--0.10 \lsun, for $n_e \lesssim 10^8$ 
cm$^{-3}$.  Although photoionization rarely produces the large electron 
temperatures, $T_e \sim 10^5$ K, implied by the intensity ratios of the 
forbidden lines, this limit on the electron density is consistent with 
the electron density derived from the flux ratios of the [Ne~V] and 
[Fe~VII] lines, $n_e \lesssim 10^7$ cm$^{-3}$.

We conclude that the [Fe~VII] and [Ne~V] emission comes from a region with 
a low electron density, $n_e \lesssim 10^7~{\rm cm^{-3}}$, and a relatively
high electron temperature $T_e > 10^5$~K. These regions have considerable
emission measures, $n_e^2 V \sim 10^{57}$ cm$^{-3}$, for
$T_e \sim 10^5$~K and $n_e \sim 10^7$ cm$^{-3}$.  The size of the 
spherical $\rm Ne^{+4}$--$\rm Fe^{+6}$ zone is $\sim$ 10 AU. Although
the hot component emits enough high energy photons to produce large
[Ne~V] and [Fe VII] emission regions, the large electron temperature 
indicates that mechanical heating also is an important energy source.
If the high ionization gas forms in a thin shell with a thickness 
of $\lesssim$ 1 AU, the shell lies $\gtrsim$ 30 AU from the central binary.

The [Mg~V] features are the last set of useful high ionization forbidden 
lines in AG Peg.
The ionization potential of $\rm Mg^{+4}$  (141~eV) is larger than 
for $\rm Ne^{+4}$ and  $\rm Fe^{+6}$. The critical density for both
$\lambda$2784 and $\lambda$2930 is $n_e = 10^8~{\rm cm^{-3}}$.
The intensity ratio, $I(\lambda$2784)/$I(\lambda$2930) = 3.4, 
is very close to the predicted value of 3.7 \citep{kaf80}.
The $I(\lambda$2784)/$I(\lambda$2417)
intensity ratio is sensitive to $n_e$ and $T_e$ \citep{kaf80}.
Our measured ratio of $I(\lambda$2784)/$I(\lambda$2417) = 10 requires
$n_e \simless 10^9~{\rm cm^{-3}}$ for $T_e > 10^5$~K.
This density is close to the critical density for $\lambda$2417.
Thus, this emission forms in denser material than the [Ne~V] and [Fe~VII] 
lines.  The emission measure for [Mg~V] suggests a shell which is factor 
of 10 or more thinner than the $\rm Ne^{+4}$--$\rm Fe^{+6}$ zone.

\subsection{Colliding Winds and Shocks}

Colliding winds have become a popular explanation for high energy phenomena 
in symbiotic and other interacting binary systems 
\citep[e.g.,][]{wal88,nus93,mur95,for95,con97,con99}.  
First developed by \citet{kwo84} and Wallerstein et al. 
\citep[1984, see also][]{wil84}, these models use the kinetic energy 
in the winds from the red giant and the hot component -- instead of 
photoionization -- to power X-rays and high ionization emission lines
observed in many systems \citep{gir87,nus89}.  The momenta in the 
two winds roughly balance in an interaction region, where shocks produce
X-ray emission and lead to the formation of highly ionized atomic
species.  The bow-shape of the interaction region also naturally produces 
emission line profiles similar to those observed in some systems.  
 
The intensity ratios of the high ionization forbidden lines in AG Peg
suggest that some of the ionized nebula in this system is mechanically 
heated \citep[see also][]{con97}.  For the hot component effective 
temperatures and bolometric luminosities observed in most symbiotic 
stars, photoionization yields much smaller electron temperatures,
$\sim$ 1--3 $\times ~ 10^4$ K, for $n_e \sim 10^5$ to $10^{10}$ 
cm$^{-3}$ \citep[e.g.,][1998]{mur94, sch97,pro96}.  
We showed in \S3.3 that photoionization can explain the observed 
line fluxes for $n_e \lesssim 10^8$ cm$^{-3}$.  Shocks from 
colliding winds are a natural mechanism to produce high 
temperature gas.  We now consider whether colliding winds 
in the system can also power the line fluxes.

Two sources of mechanical energy can potentially power the forbidden
line emission in AG Peg:
(i) an interaction region between the two stars, where material
recently lost by the hot component collides with matter recently
ejected by the red giant \citep[see][]{con97,nus93}, or
(ii) the `inner nebula' of \citet{kny91}, where material
ejected throughout the outburst compresses gas lost by the
red giant prior to the outburst.

The outer atmosphere of the red giant is a promising source of mechanically 
heated gas in AG Peg \citep[see][and references therein]{con97}.  
This region has the required density and emission measure.  
From Figure 1a of \citet{pro98}, the density in the red giant
wind is $n_e \sim 10^7$ cm$^{-3}$ at distances of $\sim$ 10 AU
from the central binary.  The density falls to $n_e \sim$ 
$3 ~ \times ~ 10^6$ cm$^{-3}$ at $\sim$ 30 AU.
The shock velocity needed to reach electron temperatures exceeding 
$10^5$ K is modest, $\sim$ 10--20 km s$^{-1}$, compared to the red giant 
wind velocity of 30--60 km s$^{-1}$ \citep{kny91}.  As envisioned in 
the \citet{wal84} picture, the colliding winds of the
red giant and the hot component might yield this shock velocity if 
the collision is oblique.  Achieving the proper forbidden line energy 
from this region, however, may be difficult.  
For a shock velocity of 10--20 km s$^{-1}$, the mechanical energy in 
the wind is $\sim$ 0.001--0.004 \lsun~for a red giant mass loss 
rate of $\sim 10^{-7} ~ \msunyr$.  This energy is small compared to 
the luminosity in a single [Ne~V] forbidden line, $\sim$ 0.05--0.10 \lsun.  
Much larger mass loss rates are ruled out by the radio data;
much larger shock velocities imply higher ionization emission lines 
such as [Fe~X] not detected on our spectra.  The wind from the 
hot component does not help this problem significantly: the mass
loss rate is now probably smaller than that of the giant \citep{ken93,
nus95} and the wind velocity of 1000 km s$^{-1}$~is much larger than 
the needed shock velocity unless we have considerably underestimated 
the electron temperature of the forbidden line region.

The inner radio nebula of \citet{kny91} is also a potential location 
for shocked gas in AG Peg.  The compressed inner shell of this nebula 
is an ideal location for shocks. The apparent gas velocity of $\sim$ 
60 km s$^{-1}$ is close to the 10--20 km s$^{-1}$ shock velocity
required for $10^5$ K gas.  The emission measure of this material, 
$n_e^2 V$ $\sim$ $10^{57}$ cm$^{-3}$ \citep{kny91}, is identical to 
our estimate for the $\rm Ne^{+4}$--$\rm Fe^{+6}$ zone.  However,
there is an important problem associating the high ionization 
forbidden lines with this region. From fits to the radio spectral 
index and flux from the inner nebula, \citet{kny91} estimate 
$n_e \approx$ $2 \times 10^4$ cm$^{-3}$ for the compressed 
shell.  This density is a factor of $\sim 10^3$ smaller than our
estimate from the [Ne~V] and [Fe~VII] line ratios.  The luminosity 
of the compressed shell is also insufficient to power the forbidden lines.
\cite{kny91} derive a rate 3 $\pm$ 3 $\times ~ 10^{-6}$ \msunyr~for
spherical mass loss, which implies a mechanical luminosity of $\sim$
0.2 \lsun~for a 20 km s$^{-1}$ shock.  The observed luminosity in the
[Ne~V] and [Fe~VII] lines exceeds this estimate by a factor of $\sim$ 2.

Both explanations for shocked gas in the AG Peg wind are sensitive to the 
adopted distance.  For a given radio flux, the mass loss rate is 
linearly proportional to the distance.  The optical emission line
luminosity depends on the square of the distance.  Better agreement
between observations and theory requires a smaller distance than our
adopted value of 800 pc. Either of our explanations for the shocked 
gas requires a distance of 400 pc or less to allow the mechanical 
energy in the wind to match the energy needed for the high ionization 
forbidden lines.  Such a small distance seems ruled out by \citet{kny91},
who prefer a 600 pc distance to reconcile the size of the inner nebula 
with proper motion data.  

We thus conclude that neither colliding wind picture for the formation of 
high ionization forbidden lines can account for the observed line fluxes 
in AG Peg.  If the hot component emits enough high energy photons to account 
for the forbidden line fluxes, mechanical heating from the colliding winds
may explain the high electron temperatures.  A detailed 
photoionization calculation which includes shock excitation would 
test this proposal.

Making more progress on the high ionization forbidden lines also 
requires new observations to constrain the geometry of the ionized nebula.
At a distance of more than 0\secpoint5 from the central binary, the
inner nebula of \citet{kny91} can be resolved with STIS on board {\it HST}.
These observations would test the notion that the forbidden emission
lines form well outside the red giant atmosphere.  Resolving the 
nebula at 30 AU, $\sim$ 0\secpoint05 from the central binary,
is more difficult but may be possible with ground-based interferometers
or future space missions. Higher spectral resolution UV or optical 
observations would yield line profiles for the high ionization lines 
and provide better constraints on the shock velocity and geometry.  

\section {Summary}

Our analysis of new optical and UV data demonstrates that the AG Peg
binary continues to evolve.  The optical data show clear evidence for
a more pronounced reflection effect, where high energy photons from 
the hot component ionize the wind and outer atmosphere of the red giant
\citep{boy66, bel70, kn83a}.  Many low
ionization optical lines, such as H~I and He~I form in this region. 
Some high ionization optical emission lines, such as He~II, also form 
in the ionized wind.  We do not have good enough UV phase coverage 
to verify that higher ionization UV lines such as C~IV or N~V are 
also produced in the ionized wind.

Detailed illumination models generally account for the observed fluxes 
of the low and moderate ionization emission lines. The H~I and He~I
fluxes imply larger mass loss rates in the red giant wind than the
fluxes of other emission lines.  Because the H~II and He~II recombination
regions in the red giant atmosphere are difficult to model accurately,
\citet[][1998]{pro96} noted that their illumination models underestimate 
the fluxes of these lines.  The illumination models thus imply a mass 
loss rate of $\sim 10^{-7}$ \msunyr~if the red giant wind has a geometric
cross-section of $\sim$ $2^2$--$3^2 R_g^2$.  This mass loss rate is
comparable to mass loss rates derived for the red giants in AG Peg and
other symbiotic stars \citep[see][]{kny91,sea92,sea93,ivi95}.

The UV data imply that the hot component continues to decline in
luminosity. Figure 7 summarizes the evolution of the optical and
bolometric magnitude $M_{bol}$ of the hot component since its eruption in
the mid 1800's.  The dashed line indicates the evolution of the
optical continuum. The symbols indicate the evolution of $M_{bol}$.
Data shown as filled symbols without boxes adopt bolometric corrections
to estimate $M_{bol}$ from optical spectra \citep{ken93};
data shown as filled symbols inside boxes use UV data from
this paper and \citet{ken93}.  As described by \citet{gal79} and
\citet{ken93}, the hot component maintained a roughly constant 
$M_{bol}$ from 1900--1980 and then began to decline.  Our new data 
suggest a factor of 2--3 decline in luminosity and a 10\%--20\% 
decline in effective temperature.  Further UV observations are 
necessary to follow the evolution of the hot component to quiescence.

The high ionization forbidden lines indicate that some portion of the
nebula is mechanically heated to temperatures of $10^5$ K or larger.
This region has a density of $\lesssim 10^7$ cm$^{-3}$ and lies at 
least 10--30 AU from the central binary.  If our estimates for the 
effective temperature and luminosity of the hot component are correct, 
the hot component emits enough high energy photons to produce the 
observed level of emission from high ionization forbidden lines.
Shocks in the red giant wind are a potential excitation mechanism 
for this gas, but the kinetic energy in the winds from the hot 
component and the red giant is smaller than the emitted energy by 
a factor of $\sim$ 10 or more. Because it has a very low density, 
a compressed shell in the inner radio nebula also seems an unlikely 
source of forbidden line emission in AG Peg.  However, this region has 
an emission measure and gas velocity close to our estimates for the 
$\rm Ne^{+4}$--$\rm Fe^{+6}$ zone.  Additional high resolution imaging 
and spectroscopic observations are needed to understand this emission.
Detailed photoionization calculations of shocked gas are necessary to
see if a combination of photoionization and mechanical heating can
explain the large amount of emission from high ionization forbidden
lines in this system.

\vskip 6ex

We acknowledge support from an archival grant from the Space
Telescope Science Institute, AR-08369-01-A, and from the 
Smithsonian Astrophysical Observatory.  This work was performed 
while D.P. held a National Research Council Research
Associateship at NASA/GSFC.  We thank J. Raymond for advice
and comments on our analysis of the high ionization forbidden
lines.

\vfill
\eject

\clearpage

%\centerline{Figure Captions}

\hskip -8ex
\epsfxsize=7.0in
\epsffile{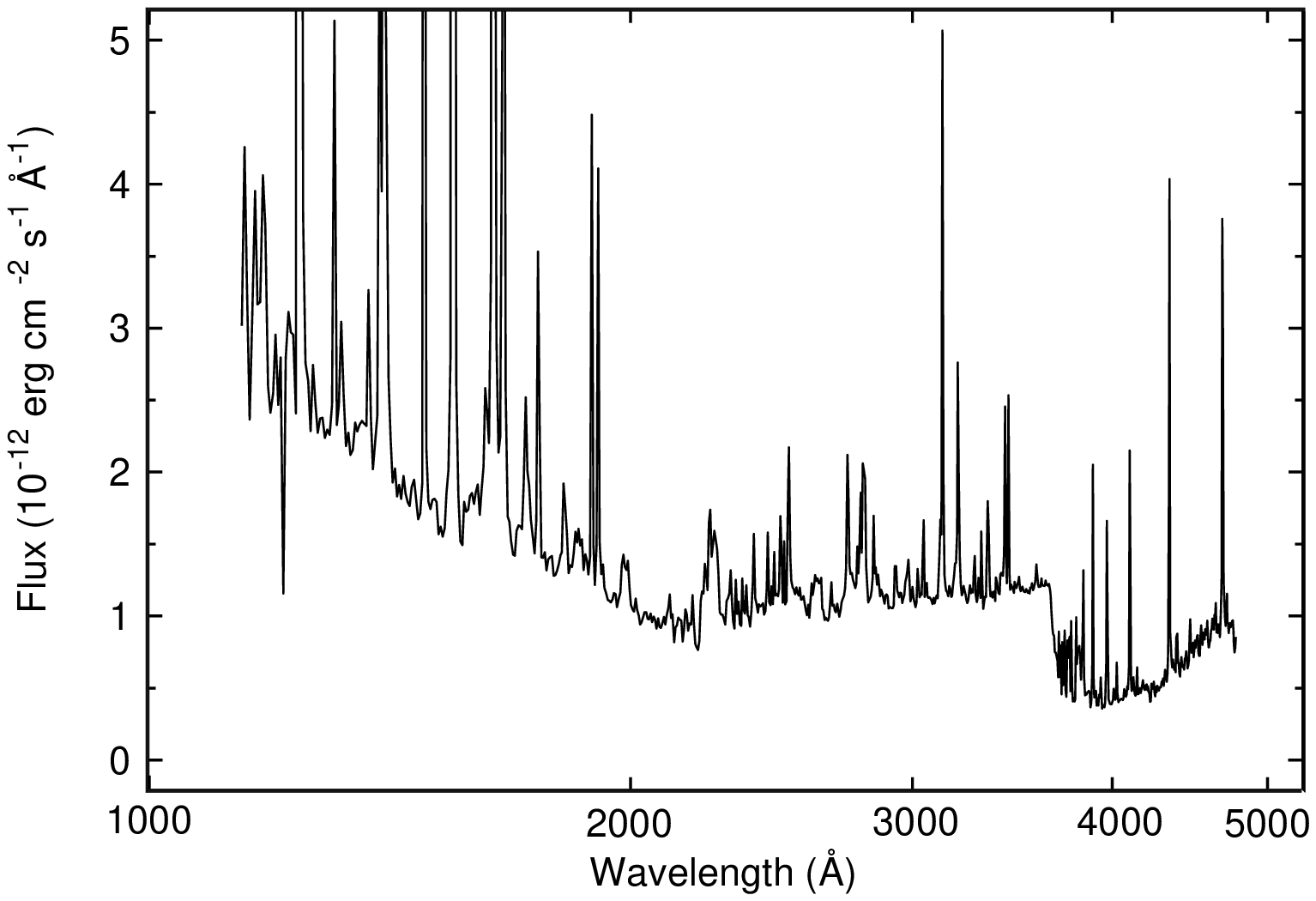}
\vskip 5ex
\figcaption[Kenyon.fig1.eps]{{\it HST} spectrum of AG Peg.
The data have not been corrected for interstellar reddening.
The continuum consists of the Rayleigh-Jeans tail of a hot blackbody
for $\lambda ~ <$ 2000 \AA. Strong nebular emission for
2000 \AA $ < ~ \lambda ~ < $ 3646 \AA, and the Wien side of
the red giant spectrum for $\lambda >$ 4200 \AA. Intense
permitted and forbidden emission lines are visible at all 
wavelengths.}

\hskip -8ex
\epsfxsize=7.0in
\epsffile{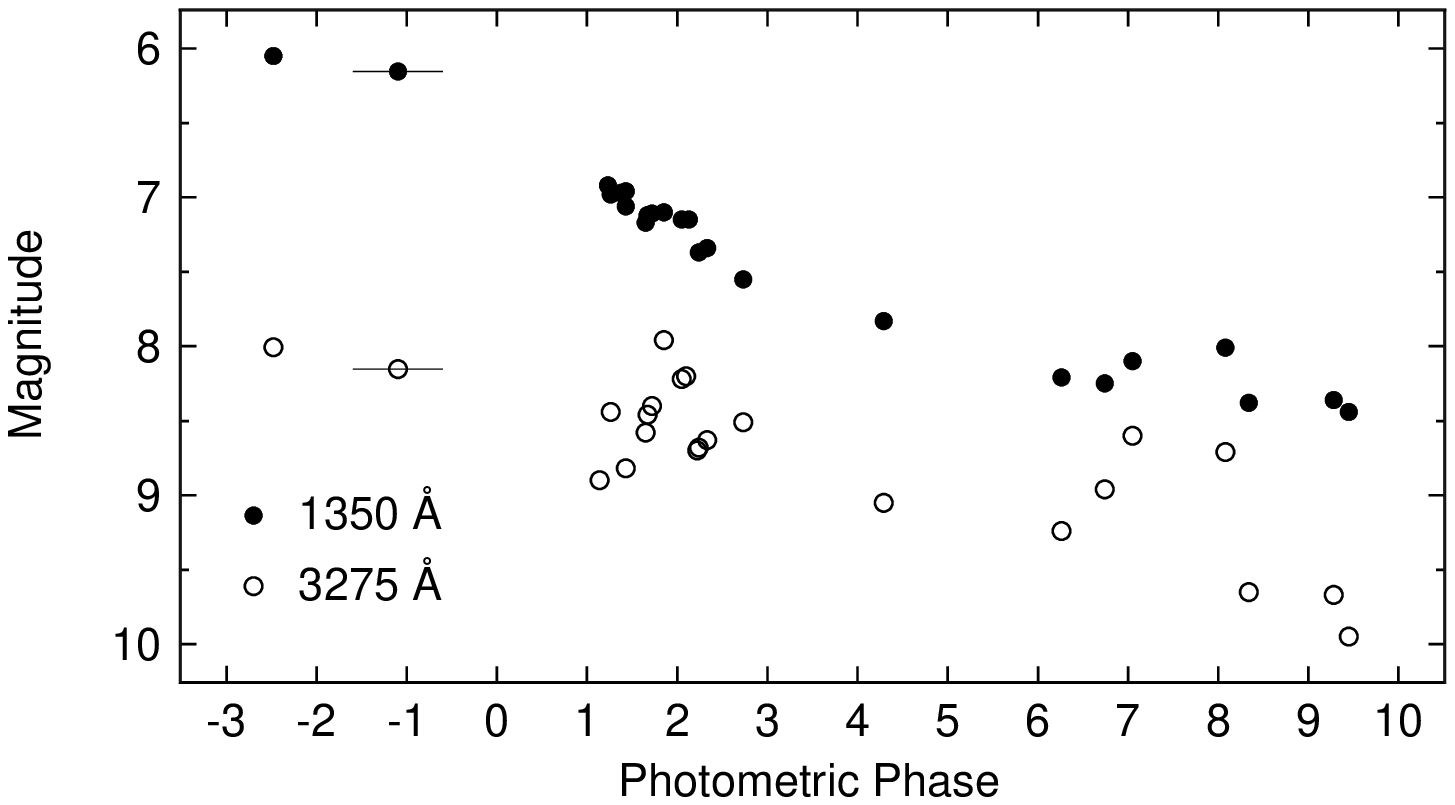}
\vskip 5ex
\figcaption[Kenyon.fig2.eps]{Variation of the UV continuum with
photometric phase.  Except for small rises at integral photometric
phases, the 1350 \AA~ and the 3275 \AA~continua decline with time.}

\hskip -10ex
\epsfxsize=7.0in
\epsffile{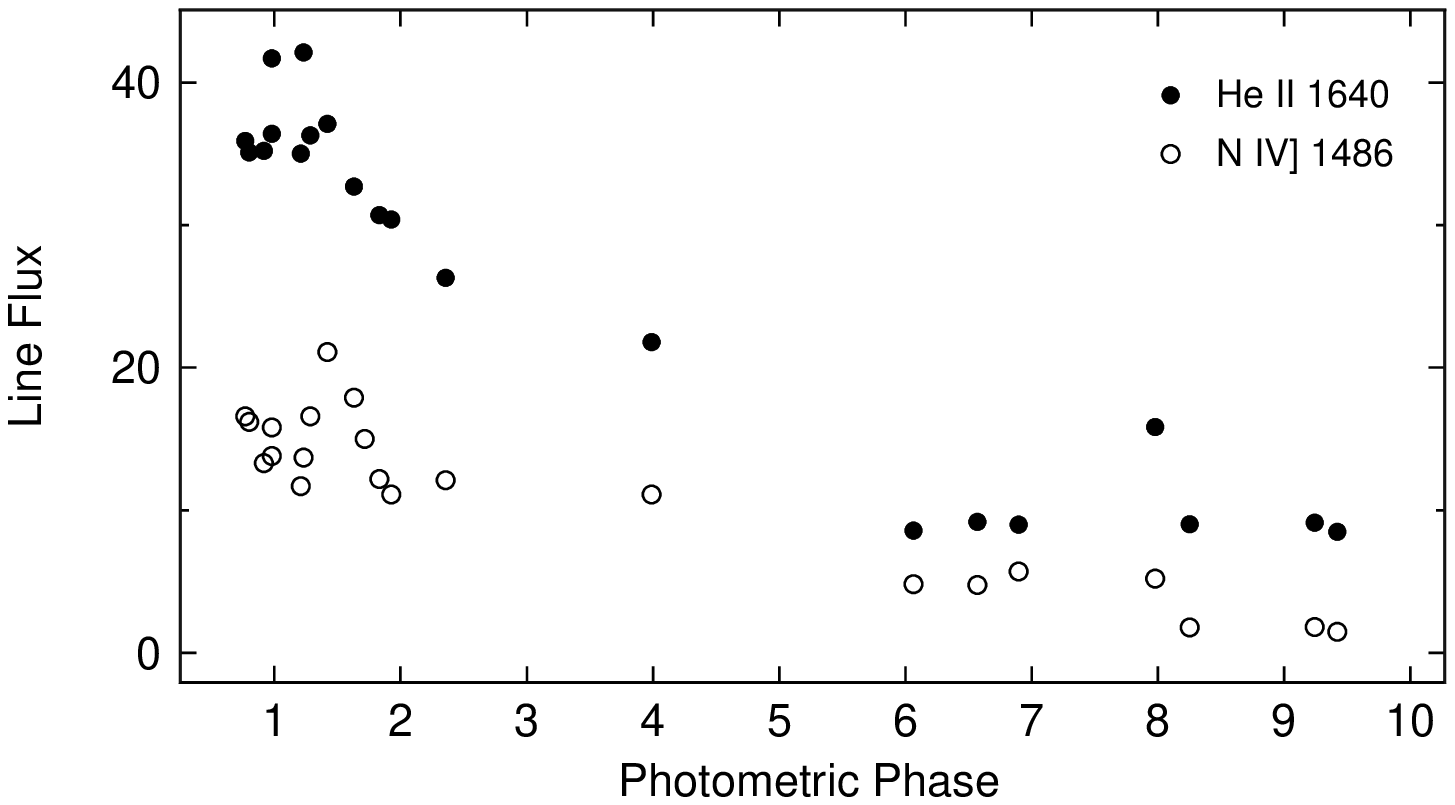}
\vskip 5ex
\figcaption[Kenyon.fig3.eps]{Variation of the He II $\lambda$1640 
and N IV] $\lambda$1486 emission lines with photometric phase.  }

\hskip -10ex
\epsfxsize=7.0in
\epsffile{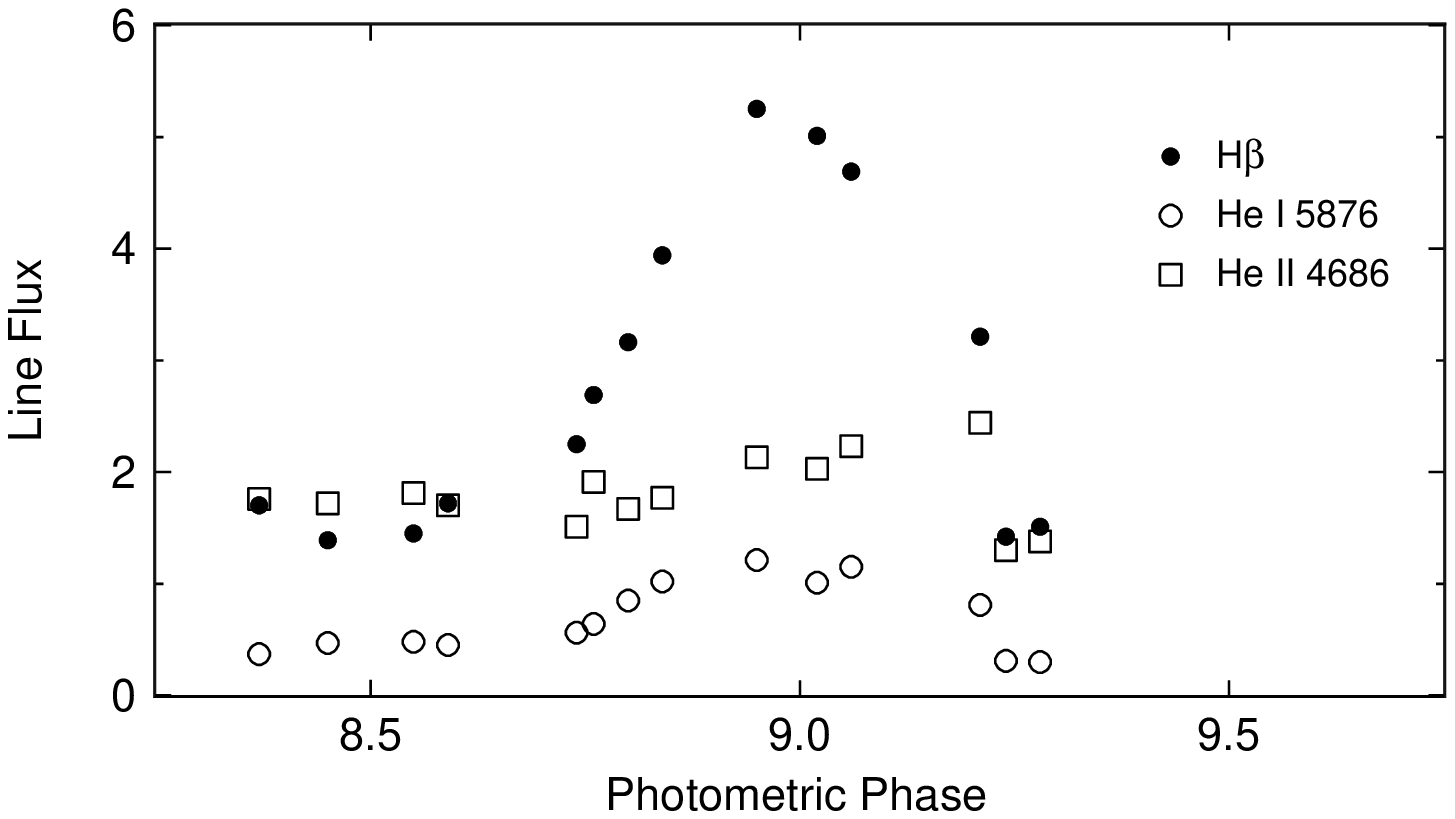}
\vskip 5ex
\figcaption[Kenyon.fig4.eps]{Variation of several optical emission
lines with photometric phase.  All of the H~I and He~I lines rise 
by factors of 2--3 at $\phi$ = 9; the He~II lines increase by less 
than 50\% at $\phi$ = 9.}

\vskip 6ex
\hskip -5ex
\epsfxsize=8.5in
\epsffile{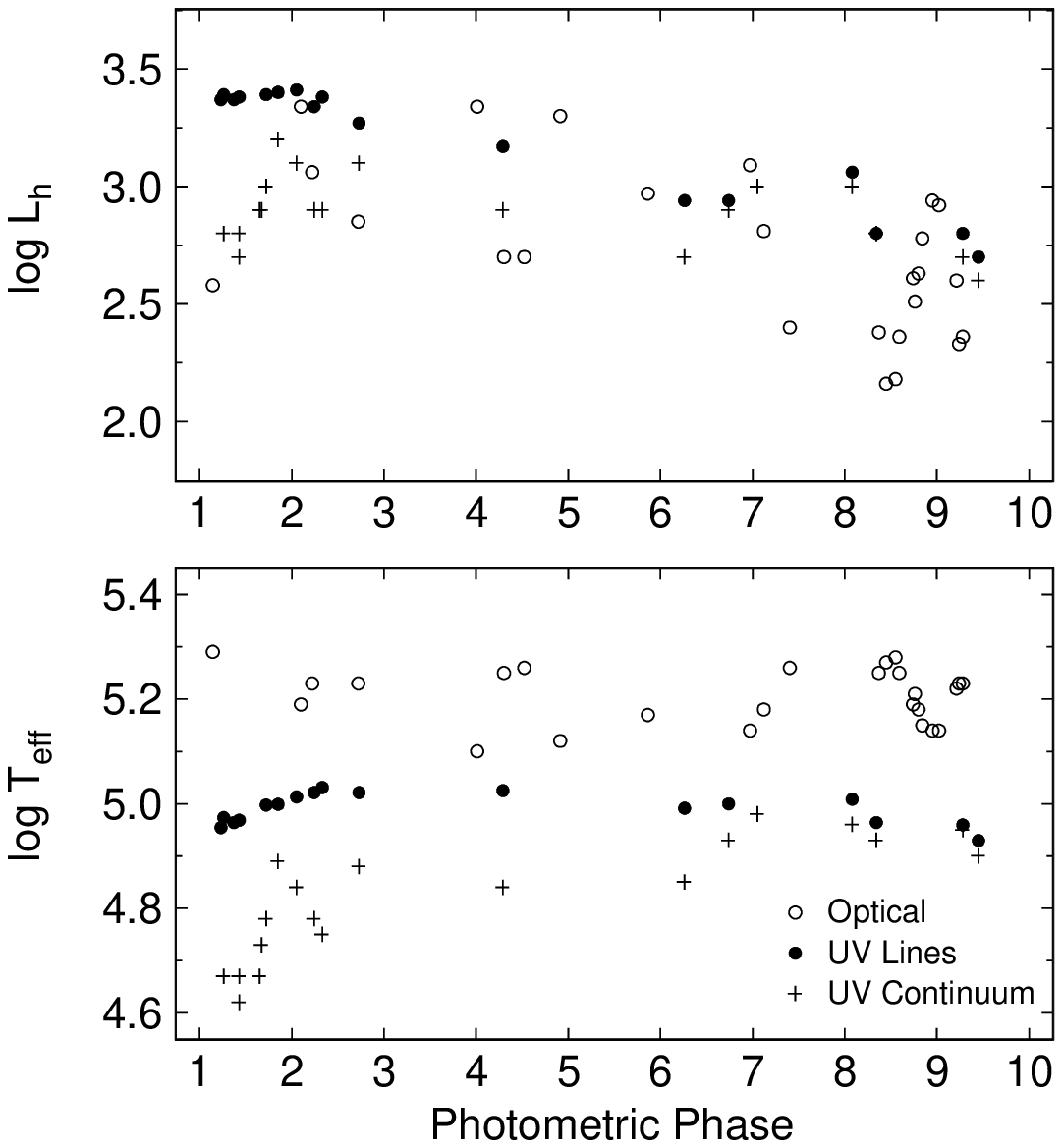}
\vskip 5ex
\figcaption[Kenyon.fig5.eps]{Variation of derived luminosity
(top panel) and effective temperature (bottom panel) of the 
hot component as a function of photometric phase.}

\hskip -7ex
\epsfxsize=6.5in
\epsffile{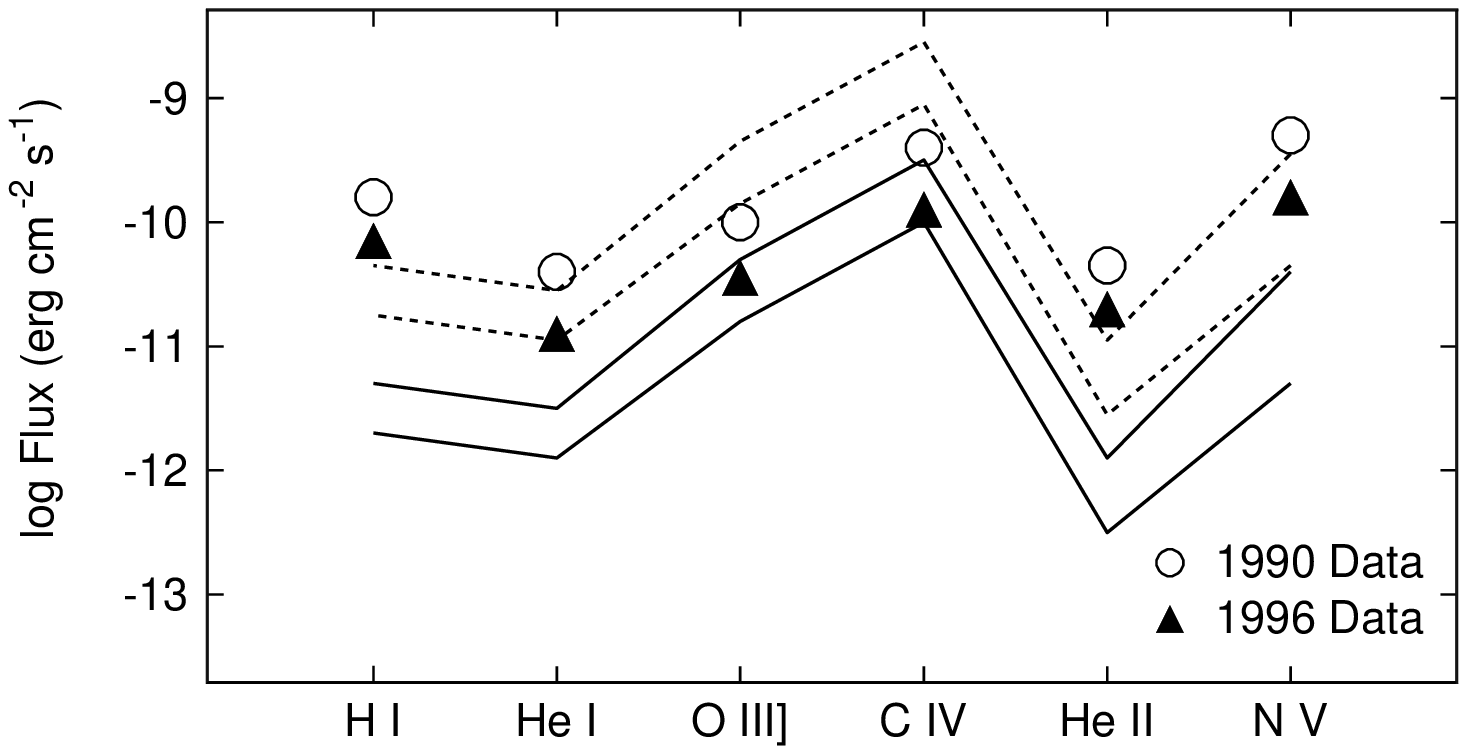}
\vskip 5ex
\figcaption[Kenyon.fig6.eps]{Comparison of illumination models
with observations of AG Peg. The legend indicates the two epochs 
of the observations.  The models assume Vogel's velocity law
for a wind with log $\beta$ = $-$2, a cross-section of 
$R_g^2$ (solid curves) or $3^2 R_g^2$ (dashed curves), and 
$\dot{M} = 10^{-6} \msunyr$ (upper curve) or
$\dot{M} = 10^{-8} \msunyr$ (lower curve).
The hot component has an effective temperature of $10^5$ K.}

\hskip -7ex
\epsfxsize=7.5in
\epsffile{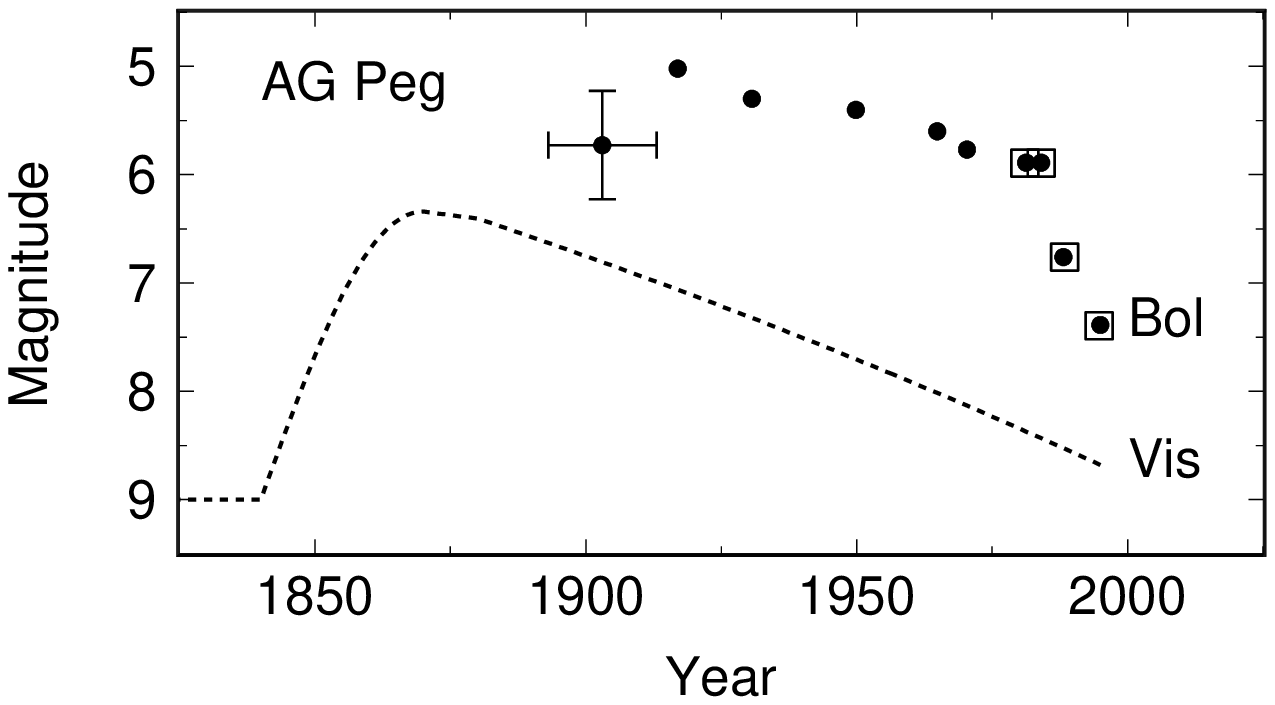}
\vskip 5ex
\figcaption[Kenyon.fig7.eps]{Variation of optical (dashed line)
and bolometric luminosity (symbols) of the hot component with
time.  The open circles are luminosities from Kenyon et al. (1993) 
using bolometric corrections estimated from optical spectra.
The cross indicates a typical error bar for these estimates.
The open boxes surrounding filled circles are luminosities 
derived from UV data in this paper and Kenyon et al. (1993).}

\clearpage

\begin{deluxetable}{lccccccc}
\tabletypesize{\small}
\tablenum{1}
\tablecaption{UV Continuum Magnitudes from HST Data}
\label{tbl-1}
\tablewidth{0pt}
\tablehead{
\colhead{JD} & \colhead{Phase} & \colhead{$\rm m_{1300}$} & 
\colhead{$\rm m_{1350}$} & \colhead{$\rm m_{1700}$} & 
\colhead{$\rm m_{2225}$} & \colhead{$\rm m_{2700}$} & 
\colhead{$\rm m_{3275}$}}
\startdata
2449304  & 8.07  & 7.95 $\pm$ 0.02 & 8.01 $\pm$ 0.02 & 8.36 $\pm$ 0.02 & 8.63 $\pm$ 0.03 & 8.89 $\pm$ 0.03 & 8.71 $\pm$ 0.02 \\
2449516  & 8.33  & 8.37 $\pm$ 0.03 & 8.38 $\pm$ 0.03 & 8.77 $\pm$ 0.03 & 9.05 $\pm$ 0.04 & 9.75 $\pm$ 0.04 & 9.65 $\pm$ 0.04  \\
2450293  & 9.27  & 8.35 $\pm$ 0.03 & 8.36 $\pm$ 0.03 & 8.80 $\pm$ 0.03 & 9.09 $\pm$ 0.04 & 9.77 $\pm$ 0.04 & 9.67 $\pm$ 0.04  \\
2450429  & 9.44  & 8.44 $\pm$ 0.03 & 8.44 $\pm$ 0.03 & 8.87 $\pm$ 0.03 & 9.15 $\pm$ 0.04 & 9.99 $\pm$ 0.04 & 9.95 $\pm$ 0.04  \\
\enddata

\tablecomments{Phases are computed according to Eq. (2).}
\end{deluxetable}

\clearpage

\begin{deluxetable}{lccccc}
\tabletypesize{\small}
\tablenum{2}
\tablecaption{Optical Continuum Magnitudes}
\label{tbl-2}
\tablewidth{0pt}
\tablehead{
\colhead{JD} & \colhead{Phase} & \colhead{$\rm m_{4400}$} & 
\colhead{$\rm m_{5550}$} & \colhead{$\rm m_{6125}$} & 
\colhead{$\rm m_{7025}$}}
\startdata
2448755 & 7.39 & 9.51$\pm$0.15 & 8.67$\pm$0.03 & 8.39$\pm$0.04 & \nodata \\
2449542 & 8.36 & 9.47$\pm$0.07 & 8.58$\pm$0.06 & 8.29$\pm$0.05 & 7.78 $\pm$ 0.05 \\
2449609 & 8.44 & 9.44$\pm$0.07 & 8.61$\pm$0.06 & 8.35$\pm$0.05 & 7.79 $\pm$ 0.05 \\
2449694 & 8.54 & 9.44$\pm$0.07 & 8.44$\pm$0.06 & 8.30$\pm$0.05 & 7.81 $\pm$ 0.05 \\
2449726 & 8.58 & 9.38$\pm$0.07 & 8.45$\pm$0.06 & 8.21$\pm$0.05 & 7.86 $\pm$ 0.05 \\
2449844 & 8.73 & 9.31$\pm$0.07 & 8.41$\pm$0.06 & 8.25$\pm$0.05 & 7.82 $\pm$ 0.05 \\
2449862 & 8.75 & 9.40$\pm$0.07 & 8.50$\pm$0.06 & 8.24$\pm$0.05 & 7.84 $\pm$ 0.05 \\
2449892 & 8.79 & 9.45$\pm$0.07 & 8.50$\pm$0.06 & 8.21$\pm$0.05 & 7.81 $\pm$ 0.05 \\
2449926 & 8.83 & 9.15$\pm$0.07 & 8.57$\pm$0.06 & 8.13$\pm$0.05 & 7.78 $\pm$ 0.05 \\
2450019 & 8.94 & 9.06$\pm$0.07 & 8.40$\pm$0.06 & 8.05$\pm$0.05 & 7.69 $\pm$ 0.05 \\
2450078 & 9.01 & 9.25$\pm$0.07 & 8.40$\pm$0.06 & 8.13$\pm$0.05 & 7.75 $\pm$ 0.05 \\
2450108 & 9.05 & 9.44$\pm$0.07 & 8.56$\pm$0.06 & 8.27$\pm$0.05 & 7.87 $\pm$ 0.05 \\
2450229 & 9.20 & 9.40$\pm$0.07 & 8.47$\pm$0.06 & 8.19$\pm$0.05 & 7.77 $\pm$ 0.05 \\
2450255 & 9.23 & 9.38$\pm$0.07 & 8.50$\pm$0.06 & 8.23$\pm$0.05 & 7.77 $\pm$ 0.05 \\
2450284 & 9.27 & 9.43$\pm$0.07 & 8.55$\pm$0.06 & 8.25$\pm$0.05 & 7.78 $\pm$ 0.05 \\
\enddata

\tablecomments{Phases are computed according to Eq. (2).}

\end{deluxetable}

\begin{deluxetable}{lcccc}
\tablecolumns{5}
\tablewidth{0pc}
\tabletypesize{\small}
\tablenum{3}
\tablecaption{HST ultraviolet emission line fluxes}
\tablehead{
\colhead{} & \multicolumn{4}{c}{JD - 2400000} \\
\colhead{Line ID} & \colhead{49304} & \colhead{49516} & 
\colhead{50293} & \colhead{50429} \\
\colhead{Photometric Phase} & 
\colhead{8.07} & \colhead{8.33} & \colhead{9.27} & \colhead{9.44}}
\startdata
N V 1240        &  20.51 & \ 6.96 & \ 6.60 & \ 3.48 \\
O I 1305        & \ 1.23 & \ 0.18 & \ 0.27 & \ 0.09 \\
O V] 1371       & \ 0.66 & \ 0.42 & \ 0.23 & \ 0.20 \\
Si IV 1394      & \ 1.49 & \ 0.29 & \ 0.35 & \ 0.13 \\
Si IV,O IV] 1403& \ 5.11 & \ 3.52 & \ 2.09 & \ 1.53 \\
S IV] 1417      & \ 0.10 & \ 0.04 & \ 0.03 & \ 0.02 \\
N IV] 1486      & \ 5.20 & \ 1.78 & \ 1.80 & \ 1.48 \\
C IV 1550       &  25.26 & \ 6.65 & \ 7.01 & \ 3.23 \\
$[$Ne V] 1574   & \ 0.12 & \ 0.10 & \ 0.11 & \ 0.12 \\
N V 1619        & \ 0.55 & \ 0.29 & \ 0.22 & \ 0.34 \\
He II 1640      &  15.83 & \ 9.03 & \ 9.12 & \ 8.49 \\
O III] 1664     & \ 4.56 & \ 1.71 & \ 1.63 & \ 1.34 \\
N IV 1719       & \ 1.29 & \ 0.71 & \ 0.61 & \ 0.41 \\
N III] 1750     & \ 1.29 & \ 0.61 & \ 0.60 & \ 0.45 \\
Si III] 1892    & \ 1.37 & \ 0.42 & \ 0.39 & \ 0.26 \\
CI III] 1908    & \ 1.20 & \ 0.68 & \ 0.65 & \ 0.53 \\
$[$Fe VII] 2015 & \ 0.05 & \ 0.03 & \ 0.04 & \ 0.03 \\
He II 2306      & \ 0.29 & \ 0.14 & \ 0.13 & \ 0.12 \\
C II] 2325      & \ 0.16 & \ 0.04 & \ 0.05 & \ 0.04 \\
He II 2385      & \ 0.34 & \ 0.17 & \ 0.16 & \ 0.14 \\
He II 2511      & \ 0.45 & \ 0.49 & \ 0.57 & \ 0.51 \\
He II 2733      & \ 0.50 & \ 0.55 & \ 0.58 & \ 0.56 \\
$[$Mg V] 2783   & \ 0.37 & \ 0.30 & \ 0.32 & \ 0.33 \\
Mg II 2793      & \ 0.74 & \ 0.19 & \ 0.19 & \ 0.11 \\
Mg II 2802      & \ 0.39 & \ 0.11 & \ 0.10 & \ 0.05 \\
O III 2832      & \ 0.18 & \ 0.09 & \ 0.10 & \ 0.09 \\
$[$Mg V] 2930   & \ 0.23 & \ 0.08 & \ 0.10 & \ 0.09 \\
$[$Ne V] 2976   & \ 0.12 & \ 0.05 & \ 0.05 & \ 0.04 \\
N III 2983      & \ 0.15 & \ 0.05 & \ 0.04 & \ 0.03 \\
O III 3023      & \ 0.12 & \ 0.06 & \ 0.08 & \ 0.05 \\
O III 3047      & \ 0.31 & \ 0.14 & \ 0.16 & \ 0.12 \\
O III 3116      & \ 0.27 & \ 0.14 & \ 0.10 & \ 0.07 \\
O III 3133      & \ 1.87 & \ 0.98 & \ 0.92 & \ 0.71 \\
He II 3203      & \ 1.51 & \ 0.87 & \ 0.92 & \ 0.87 \\
\enddata
\tablecomments{Fluxes are in units of $10^{-11} \, \rm erg ~ cm^{-2} ~ s^{-1}$} 
\tablecomments{Phases are computed according to Eq. (2).}

\end{deluxetable}

\begin{deluxetable}{lllll}
\tablecolumns{5}
\tablewidth{0pc}
\tabletypesize{\small}
\tablenum{4}
\tablecaption{HST optical emission line fluxes}
\tablehead{
\colhead{} & \multicolumn{4}{c}{JD - 2400000} \\
\colhead{Line ID} & \colhead{49304} & \colhead{49516} & 
\colhead{50293} & \colhead{50429} \\
\colhead{Photometric Phase} & \colhead{8.07} & \colhead{8.33} & 
\colhead{9.27} & \colhead{9.44}}
\startdata
O III 3299 & \ 0.05 & \ 0.04 & \ 0.03 & \ 0.03 \\
O III 3341 & \ 0.34 & \ 0.15 & \ 0.09 & \ 0.11 \\
$[$Ne V$]$ 3346 & \ 0.10 & \ 0.20 & \ 0.11 & \ 0.22 \\
$[$Ne V$]$ 3426 & \ 0.72 & \ 0.64 & \ 0.21 & \ 0.66 \\
O III 3444 & \ 0.64 & \ 0.23 & \ 0.18 & \ 0.18 \\
$[$Fe VII$]$ 3586 & \ 0.12 & \ 0.07 & \ 0.08 & \ 0.07 \\
H I 3750  & \ 0.25 & \ 0.06 & \ 0.05 & \ 0.04 \\
$[$Fe VII$]$ 3758 & \ 0.35 & \ 0.24 & \ 0.21 & \ 0.18 \\
H I 3771  & \ 0.27 & \ 0.08 & \ 0.06 & \ 0.06 \\
H I 3798  & \ 0.28 & \ 0.05 & \ 0.04 & \ 0.04 \\
H I 3835  & \ 0.35 & \ 0.21 & \ 0.19 & \ 0.21 \\
H I 3888  & \ 0.38 & \ 0.20 & \ 0.16 &  \ 0.14 \\
$[$Fe VI$]$ 3905 & \ 0.05 & \ 0.04 & \ 0.04 & \ 0.03 \\
He I 3927 & \ 0.08 & \ 0.02 & \ 0.02 & \ 0.02 \\
He I 3965 & \ 0.14 & \ 0.04 & \ 0.02 & \ 0.02 \\
H I 3970  & \ 0.78 & \ 0.21 & \ 0.14 & \ 0.14 \\
He I 4009 & \ 0.06 & \ 0.04 & \ 0.03 & \ 0.02 \\
He I 4026 & \ 0.15 & \ 0.08 & \ 0.07 & \ 0.06 \\
H I 4101  & \ 1.14 & \ 0.44 & \ 0.31 & \ 0.35 \\
He I 4121 & \ 0.05 & \ 0.02 & \ 0.02 & \ 0.02 \\
He I 4144 & \ 0.07 & \ 0.03 & \ 0.03 & \ 0.02 \\
H I 4340  & \ 2.02 & \ 0.63 & \ 0.47 & \ 0.52 \\
He I 4388 & \ 0.17 & \ 0.09 & \ 0.05 & \ 0.06 \\
He I 4471 & \ 0.15 & \ 0.07 & \ 0.05 & \ 0.05 \\
He II 4541 & \ 0.16 & \ 0.11 & \ 0.07 & \ 0.08 \\
He II 4686 & \ 2.17 & \ 1.55 & \ 1.35 & \ 1.29 \\
He I 4713 & \ 0.13 & \ 0.06 & \ 0.06 & \ 0.05 \\
\enddata
\tablecomments{Fluxes are in units of $10^{-11} \, \rm erg ~ cm^{-2} ~ s^{-1}$}
\tablecomments{Phases are computed according to Eq. (1).}

\end{deluxetable}

\begin{deluxetable}{l c c c c c c c c c c c c c c c}
\tabletypesize{\tiny}
\tablecolumns{16}
\tablewidth{7.0in}
\tablenum{5}
\tablecaption{Optical emission line fluxes in units of $10^{-11}$ erg cm$^{-2}$ s$^{-1}$.}
\tablehead{
\colhead{} & \multicolumn{15}{c}{Julian Date - 2440000} \\
\colhead{Line ID} & \colhead{8755} & \colhead{9542} & 
\colhead{9609} & \colhead{9694} & \colhead{9726} &
\colhead{9844} & \colhead{9862} & \colhead{9892} &
\colhead{9926} & \colhead{10019} & \colhead{10078} &
\colhead{10108} & \colhead{10229} & \colhead{10255} &
\colhead{10284}}
\startdata
Photometric Phase & 7.39 & 8.36 & 8.44 & 8.54 & 8.58 & 8.73 & 8.75 & 8.79 & 8.83 & 8.94 & 9.01 & 9.05 & 9.20 & 9.23 & 9.27 \\
\\
$[$Fe VII$]$ 3586  & \ 0.13   &\nodata &\nodata &\nodata &\nodata &\nodata &\nodata &\nodata &\nodata &\nodata &\nodata &\nodata &\nodata &\nodata &\nodata \\
$[$Fe VII$]$ 3759  & \ 0.34   &\nodata &\nodata &\nodata &\nodata &\nodata & \ 0.27 &\nodata &\nodata &\nodata &\nodata &\nodata &\nodata &\nodata &\nodata \\
H I 3835  & \ 0.35   & \ 0.21  &\nodata &\nodata &\nodata &\nodata & \ 0.40 & \ 0.41 & \ 0.46 & \ 0.38 & \ 0.45 & \ 0.56 & \ 0.44 & \ 0.20 & \ 0.19 \\
H I 3888  & \ 0.34   & \ 0.20  & \ 0.16  & \ 0.17 & \ 0.15 & \ 0.33 & \ 0.38 & \ 0.47 & \ 0.51 & \ 0.48 & \ 0.52 & \ 0.78 & \ 0.53 & \ 0.19 & \ 0.20 \\
H I 3970  & \ 0.36 & \ 0.22  & \ 0.21 & \ 0.20 & \ 0.23 & \ 0.34 & \ 0.40  & \ 0.77 & \ 0.81 & \ 0.98 & \ 0.89 & \ 0.85 & \ 0.55 & \ 0.21 & \ 0.21 \\
He I 4009 & \ 0.05 & \ 0.03  & \ 0.06  & \ 0.06  & \ 0.05  & \ 0.03  & \ 0.05 &\ 0.09 & \ 0.07 & \ 0.11 & \ 0.09 & \ 0.07 & \ 0.06  & \ 0.05 & \ 0.03 \\
He II 4026& \ 0.13 & \ 0.07  & \ 0.09  & \ 0.08  & \ 0.07 & \ 0.05 & \ 0.12 & \ 0.14 & \ 0.15 & \ 0.15 & \ 0.12 & \ 0.10 & \ 0.08  & \ 0.07 & \ 0.06 \\
H I 4101  & \ 0.66 & \ 0.45 & \ 0.40 & \ 0.43 & \ 0.46 & \ 0.51 & \ 0.61 & \ 0.72 & \ 0.90 & \ 1.34 & \ 1.17 & \ 1.09 & \ 0.83 & \ 0.40 & \ 0.36 \\
H I 4340  & \ 0.82 & \ 0.71  & \ 0.74  & \ 0.78  & \ 0.73  & \ 0.91  & \ 1.07  & \ 1.15  & \ 1.41  & \ 2.18  & \ 1.88  & \ 1.69 & \ 1.38 & \ 0.55 & \ 0.59 \\
He I 4388 & \ 0.11 & \ 0.09  & \ 0.07  & \ 0.08 & \ 0.07 & \ 0.08 & \ 0.11 & \ 0.11 & \ 0.13 & \ 0.22 & \ 0.12 & \ 0.22 & \ 0.10 & \ 0.06 & \ 0.05 \\
He I 4471 & \ 0.12 & \ 0.08 & \ 0.07 & \ 0.08 & \ 0.06 & \ 0.08 & \ 0.11 & \ 0.11 & \ 0.13 & \ 0.25 & \ 0.13 & \ 0.15 & \ 0.07 & \ 0.04 & \ 0.05 \\
He II 4541& \ 0.14 & \ 0.11 & \ 0.08 & \ 0.09 & \ 0.07 & \ 0.08 & \ 0.05 & \ 0.07 & \ 0.06 & \ 0.05 & \ 0.05 & \ 0.07 & \ 0.06  & \ 0.07 & \ 0.08 \\
He II 4686& \ 2.14   & \ 1.76  & \ 1.72  & \ 1.81 & \ 1.70 & \ 1.51 & \ 1.91 & \ 1.67 & \ 1.77 & \ 2.13 & \ 2.03 & \ 2.23 & \ 2.44 & \ 1.30 & \ 1.38 \\
He I 4713 & \ 0.15 & \ 0.07 & \ 0.06 & \ 0.07 & \ 0.05 & \ 0.06 & \ 0.07 & \ 0.06 & \ 0.08 & \ 0.05 & \ 0.05 & \ 0.06 & \ 0.06 & \ 0.05 & \ 0.06 \\
H I 4861  & \ 2.28 & \ 1.70 & \ 1.39 & \ 1.45 & \ 1.72 & \ 2.25 & \ 2.69 & \ 3.16 & \ 3.94 & \ 5.25 & \ 5.01 & \ 4.69 & \ 3.21 & \ 1.42 & \ 1.51 \\
He I 4922 & \ 0.17 & \ 0.15 & \ 0.11 & \ 0.12 & \ 0.10 & \ 0.18 & \ 0.25 & \ 0.32 & \ 0.37 & \ 0.47 & \ 0.51 & \ 0.29 & \ 0.25 & \ 0.20 & \ 0.15 \\
He I 5015 & \ 0.16 & \ 0.11 & \ 0.13 & \ 0.14 & \ 0.23  & \ 0.23 & \ 0.32 & \ 0.30 & \ 0.38 & \ 0.33  & \ 0.34 & \ 0.31 & \ 0.19 & \ 0.09 & \ 0.09 \\
$[$Fe VII$]$ 5721 & \ 0.15 & \ 0.16 & \ 0.15 & \ 0.16 & \ 0.15 & \ 0.13 & \ 0.20 & \ 0.12 & \ 0.12 & \ 0.21 & \ 0.22 & \ 0.22 & \ 0.22 & \ 0.13 & \ 0.12 \\
He I 5876 & \ 0.36 & \ 0.37 & \ 0.47 & \ 0.48 & \ 0.45 & \ 0.56 & \ 0.64 & \ 0.85 & \ 1.02 & \ 1.21  & \ 1.01 & \ 1.15  & \ 0.81 & \ 0.31  & \ 0.30 \\
$[$Fe VII$]$ 6087 & \ 0.21 & \ 0.20  & \ 0.15  & \ 0.16  & \ 0.15  & \ 0.13  & \ 0.17  & \ 0.14  & \ 0.11  & \ 0.19  & \ 0.18 & \ 0.17 & \ 0.20 & \ 0.20 & \ 0.14 \\
H I 6563  &\nodata & 10.20  & \ 8.53 & \ 9.17  & \ 9.02 & 11.3  & 13.00 & 12.50  & 15.70  & 21.50  & 21.80  & 22.50  & 15.40 & \ 8.70  & \ 9.25 \\
He I 6678 &\nodata & \ 0.67 & \ 0.64 & \ 0.67 & \ 0.65 & \ 0.68 & \ 0.84 & \ 0.99 & \ 1.20 & \ 1.82 & \ 1.62 & \ 1.84 & \ 0.93 & \ 0.55 & \ 0.60 \\
He I 7065 &\nodata & \ 0.51 & \ 0.44 & \ 0.47 & \ 0.43 & \ 0.41 & \ 0.49 & \ 0.51 & \ 0.61  & \ 1.13  & \ 1.20  & \ 1.09  & \ 0.49 & \ 0.44 & \ 0.37 \\
He I 7282 &\nodata & \ 0.21 & \ 0.22 & \ 0.23 & \ 0.22 & \ 0.23 & \ 0.28 & \ 0.31 & 0.32 & \ 0.37 & \ 0.39 & \ 0.41 & \ 0.25 & \ 0.19 & \ 0.16 \\
\enddata
\end{deluxetable}

\end{document}